\documentclass{aircc}
\usepackage{mathpartir}
\usepackage{hyperref}
\usepackage{listings}
\begin{document}

\title{
%{\sffamily
%J\fontsize{17}{17}\textbf{AVA MODULAR EXTENSION FOR \\ OPERATOR OVERLOADING
Java Modular Extension for \\ 
Operator Overloading
%}}
}

\author{Artem Melentyev}
\affiliation{Ural Federal University \\ \email{\url{http://artem.melentyev.me}}}

\maketitle

\begin{abstract}
The paper introduces a modular extension (plugin) for Java language compilers and Integrated Development Environments (IDE)
which adds operator overloading feature to Java language while preserving backward compatibility.

The extension use the idea of library-based language extensibility similar to SugarJ\cite{sugarj}.
But unlike most language extensions, it works directly inside the compiler and does not have any external preprocessors.
This gives much faster compilation, better language compatibility and
support of native developer tools (IDE, build tools).

The extension plugs into javac and Eclipse Java compilers 
as well as in all tools whose use the compilers such as IDEs (Netbeans, Eclipse, IntelliJ IDEA), build tools (ant, maven, gradle), etc.
No compiler, IDE, build tools modification needed. Just add a jar library to classpath and/or install a plugin to your IDE.

The paper also discuss on how to build such Java compiler extensions.

The extension source code is open on 
\url{http://amelentev.github.io/java-oo/}
\end{abstract}


\section{Introduction}
Operators are important part of a programming language.
They greatly improve readability.

For example the code
$$if~(a<b)~c[d] = -e+f*g$$
looks much cleaner than equivalent code
$$if~(a.compareTo(b)<0)~c.put(d, e.negate().add(f.multiply(g))$$
because of easy recognizable mathematical notation instead of method calls. For full list of supported operators see section \ref{typesystem}.

Generally operators defined only on primitive types.
Operator Overloading is the ability to add (and replace) operator definitions to user defined types.

Operator Overloading is a good language feature by itself,
but furthermore absence of it cause the problem:

\subsection{Operators and code reuse \label{codereuse}}
The problem is operators reduce code reuse for languages without operator overloading support (such as Java).

Suppose we have a function which does some computation on integers. Java code:
\begin{lstlisting}
int comp(int a, int b, int c) {
  return -a + b*c;
}
\end{lstlisting}
Next we realized we need to call this function on very big integers, so $int$ (and $long$) type has not enough capacity for us.
The obvious way to do this is to replace $int$ to $BigInteger$ class, which can handle arbitrary-precision integers.
But arithmetic operations don't work on non-primitive types in Java,
so we have to replace all operators to method calls:
\begin{lstlisting}
BigInteger comp(BigInteger a, BigInteger b, BigInteger c) {
  return a.negate().add(b.multiply(c));
}
\end{lstlisting}
On such simple example it doesn't look so bad, but imagine if you have to replace all operators in some very long and complex algorithm.
Moreover the code with method calls instead of operators can be reused with primitive types.

The paper addresses this problem. We present an extension which allows to supply non-primitive (user defined) types
with operator definitions, so you don't have to replace all operators just because you change the types.
Operator overloading reduce difference between primitive and reference types in Java.
So with operator overloading our $comp$ function for $BigInteger$ class looks the same as for $int$ type:
\begin{lstlisting}
BigInteger comp(BigInteger a, BigInteger b, BigInteger c) {
  return -a + b*c;
}
\end{lstlisting}

\subsection{Operator overloading and generics}
Let's go further. 
Let's try to generalize the $comp$ function from previous section.
Java language since version 5 has $generics$ support which help to generalize code.
First we create an interface for operators:
\begin{lstlisting}
interface MyNumber<TA, TM> {
  TA add(TA o);      // a + b
  TA negate();       // -a
  TA multiply(TM o); // a*b
}
\end{lstlisting}
Next, the $comp$ function:
\begin{lstlisting}
<TM, TA extends MyNumber<TA, TM>>
TA comp(TA a, TA b, TM c) {
  return -a + b*c;
}
\end{lstlisting}
First line tells to compiler that generics type \textit{TA} should implement methods from \textit{MyNumbers} interface which contain methods for our operators. And compiler will use these methods instead of operators.
Let's try to use our generalized function with geometric 2d point:
\begin{lstlisting}
class Point implements MyNumber<Point, Double> {
  double x, y;
  public Point(double x, double y) {
    this.x = x; this.y = y;
  }
  public Point add(Point o) {
    return new Point(x+o.x, y+o.y);
  }
  public Point multiply(Double o) {
    return new Point(x*o, y*o);
  }
  public Point negate() {
    return new Point(-x, -y);
  }
  public String toString() {
    return "(" + x + "," + y + ")";
  }
}
\end{lstlisting}
And $comp(new~Point(1.0,2.0), new~Point(3.0,4.0), 5.0)$ will return point (14.0, 18.0) as expected.
As you see operator overloading greatly improve code reuse and readability.

\subsection{Criticisms of operator overloading}
Operator overloading has often been criticized because it allows programmers to give operators completely different semantics depending on the types of their operands.
Java language historically doesn't support operator overloading mostly because of this reason.
The common reply to this criticism is that the same argument applies to function (method) overloading as well. 
Furthermore, even without overloading, a programmer can define a method to do something totally different from what would be expected from its name.

Also some people say that adding operator overloading to Java language will
complicate Java compiler.
The paper presents an implementation of Operator Overloading via small compiler plugin.
The patch for javac compiler has 179 modified lines, for Eclipse Java compiler - 193.

Because operator overloading allows the original programmer to change the usual semantics of an operator and to catch any subsequent programmers by surprise, it is usually considered good practice to use operator overloading with care (the same for method overloading).

\subsection{Modularity}
Adding operator overloading directly to Java compiler is only one problem
(See subprojects: forked JDK\cite{javac-oo} and Eclipse JDT\cite{eclipse.jdt-oo}).
The resulting compiler need to be installed as replacement of standard javac/ecj, updated with new versions of JDK/JDT, etc.
This complicates using of the extension, especially in big teams.

What if adding a language feature can be as easy as adding a library to a project?
The extension uses this idea of library-based language extensions, like in SugarJ\cite{sugarj}.
The Java language \textit{changes} when you compile your project with this library.

This approach has following advantages:
\begin{enumerate}
 \item Easy install and use \\
 No need to install modified compiler. Just add a library to your project.
 \item Independent of compiler changes. \\
 You do not need to modify compiler again if new version was released.
 If there are no major changes in the compiler then the plugin will work with new version just fine.
\end{enumerate}
Nonetheless, there is one disadvantage. It is harder to write such modular extensions. 
There are less ways to affect the compiler from plugin.

\section{Type system of Operator Overloading \label{typesystem}}

Our extension is semantic. It changes type system and code generation. 
Note it is impossible to implement operator overloading in Java solely on syntax level. 
We need to know operands types and it is semantic information.

Let's look at type system changes. 
The extension adds to Java type system the following type inference rules:

\subsection{Binary operators}
\begin{mathpar}
  \inferrule{e1.add(e2):T}{e1 + e2 : T} \and
  \inferrule{e1.subtract(e2):T}{e1 - e2 : T}  \and
  \inferrule{e1.multiply(e2):T}{e1 * e2 : T}  \and
  \inferrule{e1.divide(e2):T}{e1 / e2 : T} \and
  \inferrule{e1.remainder(e2):T}{e1 \% e2 : T} \and
  \inferrule{e1.and(e2):T}{e1 \& e2 : T} \and
  \inferrule{e1.or(e2):T}{e1 | e2 : T} \and
  \inferrule{e1.xor(e2):T}{e1 \wedge e2 : T} \and
  \inferrule{e1.shiftLeft(e2):T}{e1 << e2 : T} \and
  \inferrule{e1.shiftRight(e2):T}{e1 >> e2 : T}
\end{mathpar}
Here and below $e,e1,e2$ \textemdash~arbitrary expressions in Java language.
$expression : T$ means expression $expression$ has type $T$.
$e1.methodname(e2) : T$ means type (class) of expression $e1$ has a method $methodname$ what can accept $e2$ as argument,
and the method return type is $T$.

Binary operator overloading allows to write, for example, $a + b$, 
where class of $a$, has a method $add$ which can accept $b$ as argument.
Thus, $a$ and $b$ can be of type $BigInteger$ and so expression $a+b$ will be equivalent to $a.add(b)$. 

The extension doesn't change operator precedence. So $a+b*c$ will be transformed to $a.add(b.multiply(c))$ (but not $a.add(b).multiply(c)$) according to Java operator precedence rules.

\subsection{Unary operators}
\begin{mathpar}
  \inferrule{e.negate():T}{-e : T} \and \inferrule{e.not():T}{\sim e : T}
\end{mathpar}
allows to write $-a$ instead of $a.negate()$ and $\sim e$ instead of $e.not()$.

\subsection{Comparison operators}
\begin{mathpar}
  \inferrule{e1.compareTo(e2):int}{e1 < e2 : boolean \\ e1 <= e2: boolean  \\ e1 > e2: boolean \\ e1 >= e2: boolean}
\end{mathpar}
allows to write $a<b$ instead of $a.compareTo(b) < 0$. Similarly with $<=$, $>$, $>=$.

\subsection{Index operators}
\begin{mathpar}
  \inferrule{e1.get(e2):T}{e1[e2] : T} \and
  \inferrule{e1.set(e2, e3):T ~ or ~ e1.put(e2, e3):T}{e1[e2] = e3 : T}
\end{mathpar}
allows to write $list[i]$ instead of $list.get(i)$
and $list[i] = v$ instead of $list.set(i, v)$.
Thus the syntax of accessing to Java collections ($java.util.Collection, List, Map$,etc) 
become the same as the syntax of accessing to arrays.

\subsection{Assignment operator}
\begin{mathpar}
  \inferrule{e1:T1 \and T2.valueOf(T1):T2 \and var:T2}{var = e1 :T2}
\end{mathpar}
allows transforming incompatible types in assignment via static method $valueOf$.
For example $BigInteger~a = 1$ is equivalent to $BigInteger~a = BigInteger.valueOf(1)$.
This is weak version of Scala \textit{implicit conversion}\cite{itc} ($implicit$ keyword) which works only on assignments and variable declarations with initialization.
Because of this restriction assignment operator overloading doesn't cause any ambiguity (unlike in Scala).

New rules added with lowest priority, so backward compatibility is provided.
Thus any correct Java program remains correct on Java with the extension.
And some incorrect Java programs (with operator overloading) become correct on Java with the extension.

\section{Code generation}

New operator expressions are transformed (desugar) to corresponded method calls from section~\ref{typesystem}.
Comparison operator transformed to $compareTo$ method call and compare the result to 0 (e.g. $e1 <= e2$ is transformed to $e1.compareTo(e2)<=0$). Assignment operator wraps right side of the assignment to static $valueOf$ method.
So it can be considered as syntactic sugar.
But strictly speaking, it is not a syntactical extension because it doesn't change the Java syntax.

\section{Implementation}

The extension consist of 3 parts:
\begin{enumerate}
\item \textit{javac-oo-plugin} for javac compiler, Netbeans IDE and build tools (ant, maven, gradle, etc)
\item \textit{eclipse-oo-plugin} for Eclipse IDE
\item \textit{idea-oo-plugin} for IntelliJ IDEA IDE
\end{enumerate}

\subsection{Javac\label{javac}}
Development of javac extension began with experimenting on JDK 7 langtools repository\cite{javac-oo}.
An overview of javac compilation process can be found in \cite{HackerGuideJavac}.
Javac compiler has quite modular architecture and there is a module for every compilation stage in $com.sun.tools.javac.comp$ package.
The stages in order of execution are: \textit{Parse, Enter, Process, Attr, Flow, TransTypes, Lower, Generate}.
All stages except $Parse$ implemented as visitor design patterns, so it is easy to focus on something specific in abstract syntax trees (AST).
Type inference is performed in $Attr$ stage.
We cannot do operator overloading before $Attr$ stage because we need type information. 
And we cannot do it after $Attr$ stage because $Attr$ marks all overloaded operators as errors and write error messages to compiler log.
So we need to resolve types of overloaded operators inside $Attr$ stage by modifying it. 
Also $Attr$ stage uses $Resolve$ submodule to perform method and operator resolving, so we need to modify it too.
%TODO Resolve encode result in Symbol.
Next we need to desugar our constructs somewhere.
$TransTypes$ and $Lower$ are translator stages. This means they can rewrite AST.
$TransTypes$ stage translates Generic Java to conventional Java (without Generics).
$Lower$ stage desugar some high level Java constructs to low level constructs.
We need to desugar overloaded operators as soon as possible (otherwise we can miss some compiler stages), 
so we modified $TransTypes$ stage for this.
For full list of changes see difference between ``default'' and ``oo'' branches of javac-oo repository\cite{javac-oo}.

When desired functionality was achieved we began to prepare a plugin for javac while looking at difference between javac and extended javac.
Instead of modifying compiler stages we extend them by creating subclasses and override specific methods.
But original compiler stages need to be replaced by extended ones somehow.

JSR269: Pluggable annotation processing API\cite{jsr269} allows creating compiler plugins for custom annotation processing.
We do not have any annotations but we use JSR269 mechanisms to gain control during annotation processing stage \textemdash~$Process$.
Once we are in control we use $TaskListener$ of the compiler to wait until $Attr$ stage begins.
And when $Attr$ begins, we replace $Attr$, $Resolve$ and $TransTypes$ modules to our extended versions.
Not all operations described here are public, so we have to use reflection to access private members.

However, there is one interesting problem\cite{soPackagePrivate} here.
The extension needs to override a package-private method in the compiler.
Java Language Specification\cite{jls} allows access to package private fields and methods only within the same java package.
But a little known fact it also should be within the same \textit{classloader} by Java Virtual Machine specification\cite{jvmspecPP}. 
Clearly the \texttt{javac-oo-plugin.jar} (the extension) and \texttt{tools.jar} (javac compiler) have different classloaders.
As a way around this problem, the extension just injects a part of self into the compiler's classloader at run time\cite{soPackagePrivate}.

As a result we got a plain jar library, \texttt{javac-oo-plugin.jar}.
The library uses JSR269 to gain control from the compiler and then replaces some compiler modules to extended ones.
To use the javac extension just add it to classpath (\texttt{javac -cp javac-oo-plugin.jar *.java}) and enable annotation processing (on by default).

Due to changes in JDK8, there is separate fork (\texttt{javac8-oo} repository) and separate plugin for JDK8 (\texttt{javac8-oo-plugin}).

%Size of current version \texttt{javac-oo-plugin-0.4.jar} library = 19 Kilobytes.

\subsection{Netbeans IDE}

Netbeans IDE uses slightly modified javac compiler, 
so we performed some minor modification to support both Netbeans and javac in one \texttt{javac-oo-plugin.jar} library.
%After these modification the extension works fine in Netbeans. %java editor and project building.
To use the extension in Netbeans IDE you just need to activate ``Annotation processing in Editor'' in project settings.

\subsection{Eclipse Java Compiler, Eclipse Java IDE}

Eclipse Java Compiler (ecj) is completely different Java compiler.
It used by Eclipse Java IDE (Eclipse Java Developer Tools) in Java editor and project compiling.
One notable difference is that the Eclipse compiler lets you run code that didn't actually properly compile. 
If the block of code with the error is never ran, your program will run fine. 
Otherwise, it will throw an exception indicating that you tried to run code that doesn't compile.
Another difference is that the Eclipse compiler allows for incremental builds, that is, only changed and affected files compiles.

The extension development began the same way as in javac case \textemdash~by forking Eclipse JDT\cite{eclipse.jdt-oo}.
Architecture of ecj is very different from javac. Ecj is not modular.
Implementation of most ecj compilation stages is contained inside classes of abstract syntax tree (AST).
Type resolving performed in $ASTClass\#resolveType(..)$ methods.
Code generation in $ASTClass\#generateCode(..)$.
We are interested in $ArrayReference$,  $Assignment$, $BinaryExpression$, $UnaryExpression$ AST classes.

There is no visitor design pattern and there are no translators (AST rewriters).
This complicates extension much.
We can't just replace overloaded operators with method calls because of no translators.
We need to save desugared expressions and proxy many calls from nodes to these expressions.

Another problem is we can't call $\#resolveType$ twice on the same AST node.
The algorithm of ecj type resolution assumes every node is resolved only once.
Because of that there is no cache in $\#resolveType$ method. 
And if we call $resolveType$ again we will get strange name resolution error from $Scope$ due to name duplications.
Because of this problem we can't just check overloaded operators at top of $resolveType$ method
and if it is not our case then just continue below, like in javac extension.
We need to do it just before error printing.

More problems arise when we are trying to create a plugin from the extension.
We can't extend AST classes because they are hardcoded and there are no AST factory.
We need to modify AST classes directly. To do it more powerful tools can be used.

Previous version of the extension\cite{lombok-oo} uses lombok.patcher\cite{lombok.patcher} library.
It allows modifying ecj classes at runtime and reloading it.

Current version uses AspectJ load-time weaving\cite{AJLTW}. 
The patch defined as AspectJ aspects loaded at start of Eclipse IDE via Equanox weaving\cite{EquanoxW}.
AspectJ is an Aspect Oriented language based on Java.
It has much more features than specialized lombok.patcher library.
Aspect Oriented Programming fit nicely for creating plugins in non-modular environment.
It can inject code into quite specific points, like ``for all subclasses of some class, inside specific method, just before call to another specific method, do something''.

As a result we got the plugin for Eclipse IDE which uses Equanox Weaving and AspectJ for ecj compiler modification.
You just need to install the eclipse plugin to use Operator Overloading in Eclipse IDE.

\subsection{IntelliJ IDEA IDE \label{idea}}

IntelliJ IDEA uses javac or ecj for compilation. The extension support only javac for now.
Thus you need to use javac plugin (see subsection \ref{javac}) for real compilation.

But IDEA also use own complex java frontend (parser and analyzer) for Java editor.
The problem is to relax this frontend to allow operator overloading. So we need to extend type resolution only.

Like in ecj, type resolution in IDEA performed in AST nodes. 
But unlike ecj, IDEA has some private AST factory $JavaElementType$.
The extension replace in the factory some AST classes to extended ones with extra type inference.

Also IDEA has error highlighting module $HighlightVisitor$ where similar type resolution performed again.
So the extension replaces it to extended module to ignore places with correct operator overloading.

The IntelliJ IDEA extension ``Java Operator Overloading support'' is located in official IDEA repository.

% TODO: doesn't have easily method resolution functionality.

%Size: 18Kb.

\section{Related works}

Most attempts of language extending use preprocessor from extended to plain language.
But this extension works directly in the compiler, 
allowing much faster compilation, better language compatibility,
support of native developer tools (IDE, build tools, etc).

\begin{itemize}
\item jop\cite{jop}
is a preprocessor for a small subset of Java (Featherweight Java) language with operator overloading to plain Java. \\
Presented extension work for whole Java language and doesn't require preprocessing.
Also no IDE, no build tools support.
\item SugarJ\cite{sugarj}
is a Java 1.5 extension for library-based language extensibility. 
It allows changing language syntax at compile time by union syntax extensions into one grammar.
New constructs should be (finally) desugared to plain java.
Note it is impossible to implement operator overloading as SugarJ extension because it doesn't provide Java type system information.
SugarJ is built on top of Spoofax Language Workbench (SDF, Stratego, Eclipse IDE).
It uses SDF parser which is powerful but quite slow ($O(n^3)$).
It uses Eclipse Java compiler for actual compiling to bytecode. So it is a preprocessor to Java.
Works only in Eclipse based IDE (limited support, no debugging).
\item projectlombok.org uses similar ideas of compiler plugin for different needs. 
Mainly to remove Java verbosity:
generate getters, setters, toString, equals, hashCode automatically on compile time, 
type inference (val a = 1), and others.
First version of this operator overloading plugin was lombok based\cite{lombok-oo}.
\item juast\cite{juast} uses similar compiler plugin to do extra verification/bug finding.
\item javac in JDK8 has some compiler plugin support via $-Xplugin$ option\cite{Xplugin}.
Presented extension doesn't use it. It works on both JDK7 and JDK8 via JSR269.
\end{itemize}

\section{Conclusion}
The paper presented a Java modular extension (plugin) for
Operator Overloading support. 
The extension is useful for
 defining and using own operators on user defined classes
 as well as using various libraries (algebra, physics, geometry, date/time, finance) 
  with convenient mathematical syntax
  (e.g. ``$a + b$'' instead of ``$a.add(b)$'' for $java.math.BigInteger$ and $BigDecimal$).
The extension improve readability by allowing to write formulas in convenient mathematical syntax
instead of method calls and improve code reuse by reducing difference between primitive and reference types.

The extension works with:
\begin{itemize}
\item javac 7 and 8 compiler, Netbeans IDE and build tools (ant, maven, gradle, etc.)
\item Eclipse Java IDE
\item Intellij IDEA IDE
\end{itemize}
The extension source code is open and located on \\ 
\url{http://amelentev.github.io/java-oo/}

\section{Future works}
Using such ideas of compiler plugins we can add to Java language some major features like:
\begin{itemize}
\item \textit{Static extension methods} like in C\#.
\item \textit{Smart casts} like in Kotlin\cite{smartcasts}
\item \textit{Safe navigation}(?.) and \textit{elvis}(?:) operators like in Groovy\cite{safenav}
\item \textit{Off-side rule}\cite{offsiderule} syntax like in Python
\item And many more
\end{itemize}

\bibliographystyle{ieeetr}
\bibliography{ijpla}

\end{document}